\documentclass[12pt]{article}

\newcommand{\be}{\begin{equation}}
\newcommand{\ee}{\end{equation}}
\newcommand{\la}{\langle}
\newcommand{\ra}{\rangle}
\newcommand{\pt}{\partial_t}

\usepackage{graphicx}
\usepackage{times}
\usepackage{natbib}
\usepackage{fancyhdr}

\usepackage{epsfig}
\usepackage{multirow}

\begin{document}

\bibliographystyle{nature}

\renewcommand{\multirowsetup}{\centering}

\title{Reaction-diffusion processes and meta-population models
in heterogeneous networks}

\author{Vittoria Colizza$^{1,2*}$, Romualdo Pastor-Satorras$^{3}$, Alessandro Vespignani$^{1,2}$}
\maketitle

\vspace{-0.6cm}
\noindent
{\small{$^1$Complex Networks Lagrange Laboratory, Institute for Scientific Interchange (ISI), Torino, Italy}}

\noindent
{\small{$^2$School of Informatics and Department of Physics,
Indiana University, Bloomington 47406 IN}}

\noindent
{\small{$^3$Departament de F{\'i}sica i Enginyeria Nuclear, Universitat
Polit{\`e}cnica de Catalunya, Campus Nord B4, 08034 Barcelona, Spain}}

\noindent
{\small{$^*$e-mail: vcolizza@indiana.edu}}
\vspace{0.3cm}

\noindent {\bf Dynamical reaction-diffusion processes and
  meta-population models are standard modeling approaches for a wide
variety of phenomena in which local quantities~--~such as density, potential
and particles~--~diffuse and interact according to the physical laws.
  Here, we study the behavior of two basic 
  reaction-diffusion processes ($B\to A$ and $A+B \to 2B$) defined
on networks with heterogeneous topology and no limit on
  the nodes' occupation number. We investigate the effect of network
topology on the basic properties of the system's phase diagram 
and find that the network heterogeneity
sustains the reaction activity even in the limit of a vanishing
density of particles, eventually suppressing the critical point in
density driven phase transitions, whereas phase transition
and critical points, independent of the particle density, are not
altered by topological fluctuations. This work lays out
a  theoretical and
computational microscopic framework for the study of a wide range of realistic
meta-populations models and agent-based models that include the
complex features of real world networks.}\\

Reaction-diffusion (RD) processes are used to model phenomena as
diverse as chemical reactions, population evolution, epidemic
spreading and many other spatially distributed systems in which local
quantities obey physical reaction diffusion
equations\cite{Marrobook,vankampen,Murray}. At the microscopic level,
RD processes generally consist of particles (in many cases
accounting for different kinds of ``agents'', information parcels,
etc.) that diffuse in space and are subject to various reaction
processes determined by the nature of the specific problem at hand.
While fermionic RD models assume exclusion principles that limit the
number of particles on each node of the lattice, bosonic RD processes
relax these constraints and allow each node of the lattice to be
occupied by any number of particles.  The classic example is provided
by chemical reactions, in which different molecules or atoms diffuse
in space and may react whenever in close contact.  Another important
instance for bosonic RD processes is found in meta-population epidemic
models~\cite{anderson:92,may:1984,bolker:1995,lloyd:1996,%
  grenfell:1998,keeling:1995,ferguson:2003}.  In this case particles
represent people moving between different locations, such as cities or
urban areas. Individuals are divided into classes denoting their state
with respect to the modeled disease---such as infected, susceptible,
immune, etc.---and the reaction processes account for the possibility
that individuals in the same location may get in contact and change
their state according to the infection dynamics.

The above modeling approaches are based on the spatial structure of
the environment, transportation infrastructures, movement patterns,
traffic networks, etc.  The lack of accurate data on those features of
the systems were usually reflected in the use of random graphs and
regular lattices of different dimensionality as the substrate of the
RD process. This corresponds to an implicit homogeneous assumption on
the structure of the substrate, indeed used in many instances to solve
the basic equations describing the RD process. In recent years,
however, networks which trace the activities and interactions of
individuals, social patterns, transportation fluxes, and population
movements on a local and global
scale~\cite{Liljeros:2001,Schneeb:2004,Barrat:2004,Guimera:2005,Chowell:2003}
have been analyzed and found to exhibit complex features encoded in
large scale heterogeneity, self-organization and other properties
typical of complex
systems\cite{Albert:2002,Newman:2003,Doro:2003,Pastorbook}. In
particular it has been found that a wide range of societal and
technological networks exhibit a very heterogeneous topology. The
airport network among cities\cite{Barrat:2004,Guimera:2005}, the
commuting patterns in inter and intra-urban
areas\cite{Chowell:2003,transims}, and several
info-structures\cite{Pastorbook} are indeed characterized by networks
whose nodes, representing the elements of the system, have a wildly
varying degree, i.e. the number of connections to other elements.
These topological fluctuations are mathematically encoded in a
heavy-tailed degree distribution $P(k)$, defined as the probability
that any given node has degree $k$. They thus define highly
heterogeneous substrates for the RD processes that cannot be accounted
for in homogeneous or translationally invariant lattices. Analogously,
models aimed at a description of spreading processes in spatially
extended and societal systems are inevitably occurring in
meta-population networks with connectivity patterns displaying very
large fluctuations. Since connectivity fluctuations have been shown to
have a large impact on the behavior of several percolation and
fermionic systems~\cite{havlin00,newman00,Pastor:2001b}, the
investigation of their role in the case of bosonic RD processes
becomes a crucial issue for the understanding of a wide array of real
world phenomena.\\

{\bf Reaction-diffusion processes in complex networks.}

In order to investigate the effect of network heterogeneities on
the phase diagram of meta-population models and chemical reaction processes,
we consider a basic reaction scheme conserving the
number of particles that has been studied both in physics and
mathematical epidemiology, namely the reaction-diffusion process
identified by the following set of reactions~\cite{Marrobook,Cardy:1984,defreitas:2000,Pastor:2000,Kree:1989,Wijland:1998}:
\begin{eqnarray}
  B &\to& A,  \label{reaction1}\\
  B + A &\to& 2 B\,.\label{reaction2} 
\end{eqnarray}
From these reaction equations it is clear that the dynamics conserves
the total number of particles $N=N_A+N_B$, where $N_i$ is the number
of particles $i=A, B$. This process can be naturally interpreted as a
chemical reaction with an absorbing state phase
transition\cite{Marrobook,defreitas:2000,Pastor:2000}.  The same
reaction has been however used as a model problem in population
dynamics in interaction with a polluting substance\cite{Kree:1989},
and it is analogous to the classic susceptible-infected-susceptible
(SIS) model for epidemic spreading~\cite{anderson:92,Wijland:1998}.
In the process described by eqs.~(\ref{reaction1})-(\ref{reaction2}) the dynamics is exclusively due to $B$ particles, that
we identify as {\em active} particles, since $A$ particles cannot
generate spontaneously $B$ particles.  We consider the particles
diffusing on a heterogeneous network with $V$ nodes having a degree
distribution $P(k)$ characterized by the first and second moments $\langle
k\rangle$ and $\langle k^2\rangle$, respectively.  Reaction processes take place inside
the network's nodes only, where each node $i$ stores a number $a_i$ of
$A$ particles and $b_i$ of $B$ particles (see Figure 1).
The occupation numbers $a_i$ and $b_i$ can assume any integer value,
including $a_i=b_i=0$, that is, void nodes with no particles.  For the
sake of simplicity we assume that $B$ particles diffuse with unitary
time rate $D_B=1$ along one of the links departing from the node in
which they are at a given time. This implies that at each time step a
particle sitting on a node with degree $k$ will jump into one of its
nearest neighbor with probability $1/k$. The results obtained in the
following may be recovered for any diffusion rate $D_B$, at the
expense of a more complicate mathematical treatment that will be
reported elsewhere.  In the case of $A$ particles, we consider two
different situations corresponding to a unitary ($D_A=1$) and a null
($D_A=0$) diffusion rate, respectively.  While the first case is used
in epidemic models  which consider all individuals diffusing with the same
rate, the case $D_A=0$ is used in self-organized critical systems and
specific absorbing phase transitions coupled with non-diffusive
fields\cite{Wijland:1998,defreitas:2000,Pastor:2000}.  
When $D_A=0$, the diffusion of $A$ particles in the network 
occurs only through an effective
process mediated by the reaction with $B$ particles. Indeed any $A$
particle may become a $B$ particle following the reaction process and
diffuse in the network until the reaction $B\to A$ occurs. This process
is thus equivalent to an effective diffusion of $A$ particles in the
network.

Before the diffusion process, the $a_i$ and $b_i$ particles stored in
the same node react according to eqs.~(\ref{reaction1}) and
(\ref{reaction2}).  In each node $i$ the spontaneous process $B\to A$
simply consists in turning each $b_i$ particle into an $a_i$ particle
with rate $\mu$.  We consider two general forms for the $B+A\to 2B$
process. In {\em type I} reaction we consider that each $a_i$ may
react with all the $b_i$ particles in the same node, each reaction
occurring with rate $\beta$. In {\em type II} reaction we consider
instead that each particle has a finite number of contacts with other
particles. In this case the reaction rate has to be rescaled by the
total number of particles present in the node, i.e. $\beta/\rho_i$,
where $\rho_i= a_i+b_i$ is the total number of particles in the node.
This second process corresponds to what we usually observe in epidemic
processes, where there is a population dependence of the reaction rate
since individuals generally meet with a definite number of other
individuals. In regular lattices and within the homogeneous mixing
(mean-field) hypothesis, both type of processes exhibit a phase
transition from an active phase (with an everlasting activity of B
particles) to an absorbing phase (devoid of B particles), which in
epidemic modeling correspond to the infected and healthy states,
respectively. In the type I reaction the relevant parameter is
represented by the average density of particles $\rho=N/V$ and the
transition occurs at the threshold value $\rho=\rho_c=\mu/\beta$~\cite{defreitas:2000,Pastor:2000,Wijland:1998}. In
the type II processes the transition point is found whenever
$\beta/\mu >1$. This second case is analogous to the classical
epidemic threshold result and determines the lack or existence of an
endemic state with a finite density of infected individuals (in this
representation corresponding to the $B$ particles)\cite{Murray,anderson:92}.

In order to take into account the topological fluctuations of the
networks we have to explicitly consider the presence of nodes with
very different degree $k$. A convenient representation of the system
is therefore provided by the quantities
\begin{equation}
\rho_{A,k}=\frac{1}{v_k}\sum_{i|k_i=k} a_i\,, 
~~~~~~~~~~\rho_{B,k}=\frac{1}{v_k}\sum_{i|k_i=k} b_i\,,
\end{equation}
where $v_k$ is the number of nodes with degree $k$ and the sums run
over all nodes $i$ having degree $k_i$ equal to $k$. These two
quantities express the average number of $A$ and $B$ particles in
nodes with degree $k$. Analogously, $\rho_k=\rho_{A,k}+\rho_{B,k}$
represents the average number of particles in nodes with degree $k$.
The average density of $A$ and $B$ particles in the network is given
by $\rho_{A}=\sum_k P(k)\rho_{A,k}$ and $\rho_{B}=\sum_k
P(k)\rho_{B,k}$, respectively. Finally, by definition it follows that
$\rho=\rho_{A}+\rho_{B}$.  These quantities allow to express the RD
process occurring on a heterogeneous network in terms of a set of rate
equations describing the time evolution of the quantities
$\rho_{A,k}(t)$ and $\rho_{B,k}(t)$ for each degree class $k$, as
reported in the Materials and Methods section.  The equations depend
on the reaction kernel $\Gamma_k$ that yields the number of $B$
particles generated per unit time by the reaction processes taking
place in nodes of a given degree class $k$. In uncorrelated networks
the resulting equations can be solved in the stationary limit, thus
providing information on the phase diagram of the processes.\\

{\bf Phase diagram and critical threshold.}

Let us first consider the type I reaction processes in uncorrelated
networks. In this case we have that $\Gamma_k=\rho_{A,k}\rho_{B,k}$
obtaining in the stationary state \be \rho_{B,k}=\frac{k}{\la k\ra}
\left[(1-\mu)\rho_B+\beta \Gamma \right],
\label{eq:IrhokB}
\ee where $\Gamma=\sum_k P(k)\Gamma_k$. This equation readily states
that the density of active particles is increasing in nodes with
increasing degree $k$. This effect is mainly due to the diffusion
process, that brings a large number of particles to well connected
nodes, reflecting thus the impact of the network topological
fluctuations on the particle density behavior (see the online supplementary
information). In order to study the
phase diagram we have to find the condition for which  a
solution $\rho_B\neq0$ of the set of equations for $\rho_{A,k}(t)$ and
$\rho_{B,k}(t)$ is allowed.  If $D_A=0$, i.e. for the case of non diffusing $A$
particles, the density of $A$ particles is independent of the node
degree and is given by $\rho_{A,k}=\rho_A=\mu/\beta$. In view of the
conservation of the number of particles this result readily implies
that $\rho_B=\rho-\frac{\mu}{\beta}$ and therefore the presence of a
phase transition from an absorbing phase to an active state at a
critical value of the total density of particles
$\rho_c=\frac{\mu}{\beta}$.  A very different picture is obtained when
$A$ particles are also allowed to diffuse. In this case, for $D_A=1$,
the stationary density of $A$ particles is given by
\begin{equation}
 \rho_{A,k}=\frac{k}{\la k\ra}
(\rho_A+\mu\rho_B-\beta \Gamma)\,.
\label{eq:IrhokA}
\end{equation}
The system of equations can be solved by imposing a self-consistent
condition for the quantity $\Gamma$ (the details of the calculation
are reported in the online supplementary information) and the non-trivial
solution $\rho_B>0$ is obtained only if the total density of particles
satisfies the condition $\rho>\rho_c$, with
\begin{equation}
\rho_c=\frac{\la k\ra^2}{\la k^2\ra} \frac{\mu}{\beta}\,.
\label{eq:I_rhoc_DA1}
\end{equation}
This result implies that, if $A$ particles can diffuse, topological
fluctuations affect the critical value of the transition.  Networks
characterized by heterogeneous connectivity patterns display large
degree fluctuations so that $\la k^2\ra \gg \la k\ra^2$. In the infinite
size limit $V\to\infty$ we have $\la k^2\ra\to\infty$ and thus
eq.~(\ref{eq:I_rhoc_DA1}) yields a critical value $\rho_c=0$, showing
that the topological fluctuations of the network suppress the phase
transition in the thermodynamic limit.  This is a very relevant result
that, analogously to those concerning percolation~\cite{havlin00,newman00} and
standard epidemic processes~\cite{Pastor:2001b}, indicates that
physical and dynamical processes taking place on scale-free and
heavy-tailed networks behave very differently with respect to the same
processes occurring on homogeneous networks.

In Figure 2 we provide support to this theoretical
picture by reporting the results obtained from Monte Carlo simulations
of RD processes of type I on uncorrelated networks with given
scale-free degree distribution $P(k) \sim k^{-\gamma}$.  The simulations use a
single particle modeling strategy in which each individual particle is
tracked in time.  The system evolves following a stochastic
microscopic dynamics and at each time step it is possible to record
average quantities, such as e.g.  the density of active particles
$\rho_B(t)$.  In addition, given the stochastic nature of the dynamics,
the experiment can be repeated with different realizations of the
noise, different underlying graphs, and different initial conditions.
This approach is equivalent to the real evolution of the RD process in
the generated networks and can be used to validate the theoretical
results obtained in the analytical approach.  The top panel of Figure
2 shows the phase transitions observed in the two cases, whether A
particles diffuse or not. If $D_A=0$ the process undergoes a phase
transition at $\rho_c=\mu/\beta=2$, regardless of the difference in the level
of heterogeneity as provided by different values of the power-law
exponent $\gamma$ of the degree distribution $P(k)$ at fixed network size
(here $V=10^4$, $\gamma=3$ and $\gamma=2.5$) or by different sizes $V$ at fixed
$\gamma$ ($\gamma=2.5$, $V=10^4$ and $V=10^5$). In case A particles diffuse,
instead, the transition occurs at critical values $\rho_c<\mu/\beta$, with
$\rho_c\to 0$ for decreasing ratios $\la k\ra^2/\la k^2\ra$ as observed for
increasing sizes $V$ at fixed $\gamma$ (curves for $\gamma=2.5$ with $V$ from $10^3$
to $10^5$ nodes are shown), in agreement with the analytical result
of eq.~(\ref{eq:I_rhoc_DA1}).  Bottom panels show the difference in
the behavior of $\rho_{A,k}$ as a function of the degree $k$: a flat
spectrum is obtained when $D_A=0$ and a linear dependence in $k$ when
$D_A=1$.

A different scenario emerges when considering type II processes. In
this case the reaction kernel is $\Gamma_k=\rho_{A,k}\rho_{B,k}/\rho_k$; i.e. in
each node $A$ particles will participate in a reaction event with a
rate proportional to the relative density of $B$ particles. While the
set of equations for $\rho_{B,k}$ has the same form of
eq.~(\ref{eq:IrhokB}), the stationary condition for $\rho_{A,k}$ yields
solutions that depend on $k$ for both diffusive and non-diffusive $A$
particles. In particular, in both cases we have that $\rho_B>0$ if the
condition $\beta/\mu>1$ is satisfied (see the online supplementary
information). This result recovers the usual threshold condition which
depends only on the reaction rates and is not affected by changes in
the total density of particles $\rho$.  Also for type II processes we
performed extensive Monte-Carlo simulations considering uncorrelated
scale-free networks with a heavy tailed degree distribution.
Figure 3 reports the results obtained in the two cases
$D_A=0$ and $D_A=1$, with different underlying network topologies.
Changes in the number of nodes and in the exponent $\gamma$ assumed for the
degree distribution do not affect the phase transition. The critical
value depends exclusively on the process rates, despite the observed
linear behavior of $\rho_{A,k}$ and $\rho_{B,k}$ bears memory of the
heterogeneity of the underlying network.\\

{\bf Discussion and comparison with realistic models.}

The different phase diagrams obtained in type I and II processes can
be understood qualitatively in terms of the following argument. In
type I processes, whatever the parameters $\beta$ and $\mu$, there exists a
value of the local density large enough to keep the system in an
active state by sustaining the creation of $B$ particles in the right
amount. Large topological fluctuations imply the existence of high
degree nodes with a high density of particles and therefore a high
number of generated $B$ particles.  This implies that in the
thermodynamic limit there is always a node (with a virtually infinite
degree) with enough particles to keep alive the process even for a
vanishing average density of particles, leading to the suppression of
the phase transition. The crucial ingredient of this mechanism is
given by the diffusion process that allows high degree nodes to have a
number of particles that is proportional to their degree. This is
confirmed by the case in which $A$ particles do not diffuse. In this
case high degree nodes do not accumulate enough particles and the
usual threshold effect is recovered.  In type II processes, on the
other hand, the reaction activity in each node is rescaled by the
local density $\rho_i$ and it is therefore the same in all nodes,
regardless of the local population.  In this case, the generation of
$B$ particles is homogeneous across nodes of different degrees and
therefore the presence of an active state depends only on the balance
between the reaction rates $\beta$ and $\mu$.

These results let emerge a basic framework for the microscopic
(mechanistic in the epidemic terminology) description of
meta-population epidemic models. The type I and II processes
correspond to the two limits of transmissibility independent or
inversely proportional to the population size, respectively. In
addition, realistic meta-population models have heterogeneous
diffusion probabilities due to the traveling pattern and fixed
population sizes according to data. Despite these extra complications,
the basic reaction-diffusion framework studied here provides a simple
qualitative picture of the realistic models.  In Figure 4
we report the two types of processes studied here and compare them
with the results from a realistic compartmental SIS meta-population
model considering 500 urban areas in the Unites States and including
the actual data of the air traveling flows among those urban
areas~\cite{iata,census}. The network is defined by nodes representing
each urban area together with its population and edges representing air travel
fluxes along which individuals diffuse, coupling the epidemic spreading
in different urban areas (see Ref.\cite{Longini,Geisel,Colizzapnas}
for a detailed definition of the model).  The type I process is
compared with a model whose transmissibility is independent of the
population size and the type II process is compared with the usual
epidemic spreading with trasmissibility scaling proportionally to the
population size (see Ref.~\cite{may:1984}).  Figure 4 shows
in the four cases the reaction activity occurring on each network
node, i.e.  the creation of B particles in the reaction-diffusion
process and newly infected individuals in the realistic epidemic model
normalized to the local population.  Increasing values of the reaction
activity correspond to colors ranging from  yellow (low activity)
to  red (high activity). In type I processes the reaction activity
is linearly increasing with the population of the nodes, thus showing
high activity (red) concentrated in largely populated nodes
(represented with a larger size).  The homogeneity in the generation
of B particles in type II processes is evident: all nodes display the
same color and thus experience the same level of activity, regardless
of the local density. Strikingly, despite the various complications
and elements of realism introduced in the data-driven meta-population
model, its qualitative behavior is in very good agreement  with the results obtained for the microscopic
reaction-diffusion processes in both
transmissibility limits.

In summary, the microscopic RD framework introduced here is able to
provide a general theoretical understanding of the behavior of more
realistic meta-population epidemic models.  Furthermore, the presented
approach can be extended to include the various sources of
heterogeneity---such as degree
correlations\cite{Pastor:2001,Newman:2002}, heterogeneous diffusion
probabilities and their non-linear relations with the connectivity
pattern---needed in order to provide a detailed analysis of realistic
processes.\\

{\bf Materials and Methods}

{\em Reaction-diffusion equation}.  In order to take fully into
account degree fluctuations in an analytical description of the RD
processes, we have to relax the homogeneity assumption and allow for
degree fluctuations by introducing the relative densities
$\rho_{B,k}(t)$, $\rho_{A,k}(t)$, and $\rho_{k}(t)$. The dynamical
reaction rate equations for $B$ particles in any given degree class
can thus be written as
\begin{equation}
  \pt\rho_{B,k}=-\rho_{B,k}+k\sum_{k'}P(k'|k)\frac{1}{k'}\left[(1-\mu)
    \rho_{B,k'}+\beta \Gamma_{k'}\right]\,, 
\end{equation}
where $P(k'|k)$ represents the conditional probability that a vertex
of degree $k$ is connected to a vertex of degree
$k'$\cite{Pastor:2001}.  The various terms of the equations are
obtained by considering that at each time step the particles present
on a node of degree $k$ first react and then diffuse away from the
node with unitary diffusion rate accounted by the term $-\rho_{B,k}$. The
positive contribution for the particle density is obtained by summing
the contribution of all particles diffusing in nodes of degree $k$
from their neighbors of any degree $k'$, including the new particles
generated by the reaction term $\Gamma_k$.  In the case of uncorrelated
networks the conditional probability $P(k'|k)$ that any given edge
points to a vertex with $k'$ edges is independent of $k$ and equal to
$k'P(k')/\langle k \rangle$~\cite{Doro:2003,Pastor:2001}, so that the reaction
rate equations read as \be \pt\rho_{B,k}=-\rho_{B,k}+\frac{k}{\la k\ra}
\left[(1-\mu)\rho_B+\beta \Gamma \right], \ee where $\rho_B=\sum_k P(k) \rho_{B,k}$ and
$\Gamma=\sum_k P(k)\Gamma_k$. In the case for $\rho_{A,k}(t)$ we have to distinguish
whether if $A$ particles diffuse or not.  If $D_A=1$, we obtain a set
of equations analogous to those for $\rho_{B,k}$ that read as \be
\pt\rho_{A,k}= -\rho_{A,k}+ \frac{k}{\la k\ra} (\rho_A+\mu\rho_B-\beta \Gamma)\,,
\label{eq:IIrhokA}
\ee where $\rho_A=\sum_k P(k) \rho_{A,k}$.  In the case of
non-diffusive $A$ particles ($D_A=0)$ the equations reduce to:
\begin{equation}
\pt \rho_{A,k} = \mu \rho_{B,k}-\beta\Gamma_k.
\label{eq:IIrhokAnd}
\end{equation}
The phase diagram for the various cases and the conditions for
$\rho_B>0$ are obtained by imposing the stationary state defined by
$\pt \rho_{A,k}=0$ and $\pt\rho_{B,k}=0$, with the additional
constraint that $\rho=\rho_A+\rho_B$, i.e. the number of particles is
conserved. We are therefore led to a simple set of algebraic equations
whose explicit solution is reported in the online supplementary
information.

{\em Monte-Carlo simulations}. The uncorrelated networks considered
have been generated with the \emph{uncorrelated configuration
  model}~\cite{Catanzaro:2005}, based on the
\emph{Molloy-Reed}~\cite{MR}  algorithm with an additional
constraint on the possible maximum value of the degree in order to
avoid inherent structural correlations.  The algorithm is defined as
follows.  Each node $i$ is assigned a degree $k_i$ obtained from a
given degree sequence $P(k)$ (in our case $P(k)\sim k^{-\gamma}$ with
$\gamma=3$ and $\gamma=2.5$) subject to the restriction $k_i<V^{1/2}$.
Links are then drawn to randomly connect pairs of nodes, respecting
their degree and avoiding self-loops and multiple edges.  Sizes of
$V=10^3$, $V=10^4$ and $V=10^5$ nodes have been considered.  Initial conditions
are generated by randomly placing $V\rho_A(0)$ particles $A$ and $V
\rho_B(0)$ particles $B$, corresponding to a particle density
$\rho=\rho_A(0)+\rho_B(0)$.  The results are independent of the
particular initial ratio $\rho_A(0)/\rho_B(0)$, apart from very early
time transients.  The dynamics proceeds in parallel and considers
discrete time steps representing the unitary time scale $\tau$ of the
process. The reaction and diffusion rates are therefore converted into
probabilities and at each time step, the system is updated according
to the following rules.  a) Reaction processes: \emph{i)} On each
lattice site, each $B$ particle is turned into an $A$ particle with
probability $\mu\tau$. \emph{ii)} At the same time, each $A$ particle
becomes a $B$ particle with probability determined by the type of
reaction process.  b) After all nodes have been updated for the
reaction, we perform the diffusion: on each lattice site, each $B$
particle moves into a randomly chosen nearest neighbor site; the same process occurs for $A$ particles if $D_A=1$.  The
simulation details of the reaction process represented by
eq.~(\ref{reaction2}) depend on the kernel considered.  In type I
processes each $A$ particle in a given node $i$ becomes a $B$ particle
with probability $1-(1-\beta\tau)^{b_i}$, where ${b_i}$ is the total
number of $B$ particles in that node. This corresponds to the average
probability for an $A$ particle of being involved in the
reaction~(\ref{reaction2}) with any of the $B$ particles present on
the same site.  In type II processes the reaction process is simulated
by turning each A particle into a B particle with probability
$1-(1-\frac{\beta}{\rho_i}\tau)^{b_i}$ where $\rho_i$ is the total
number of particles in the node $i$.  This term accounts for the
average probability that an A particle will get in contact with a B
particle present in the node, given that the possible number of
contacts is rescaled by the population $\rho_i$ of the node.  The term
$\beta/\rho_i$ therefore represents the normalized transmission rate
of the process.

\bigskip
\bigskip
\noindent
Correspondence and requests for material should be addressed to V.C.
(vcolizza@indiana.edu) or A.V. (alexv@indiana.edu).

\bigskip
\noindent
{\bf Acknowledgments}
A.V. is partially supported by the NSF award IIS-0513650. R. P.-S.
acknowledges financial support from the Spanish MEC (FEDER), under
project No.  FIS2004-05923-C02-01 and additional support from the
DURSI, Generalitat de Catalunya (Spain).

\newpage
\lhead{\emph{FIGURES}}

\noindent
{\Large{\bf Figure captions}}
\vspace{1cm}

\noindent
{\bf Figure 1.} {\bf Bosonic reaction-diffusion systems in heterogeneous
    networks.} Schematic representation of RD processes in
  heterogeneous complex networks when the multiple occupancy of nodes
  is allowed. Particles $A$ and $B$ can diffuse in the network and,
  inside each node, undergo the reaction processes described by
  eqs.~(\ref{reaction1})--(\ref{reaction2}). Each node $i$  stores
  $\rho_i=a_i+b_i$ particles, where the occupation numbers $a_i$ and
  $b_i$ can assume any integer value, including zero.\\

\noindent
{\bf Figure 2.} {\bf Phase diagram and stationary densities for type I
    processes.} Top panel: Phase transitions in type I processes for
  diffusing and non-diffusing $A$ particles. If $D_A=0$ the transition
  occurs at the critical value of the density $\rho_c=\mu/\beta=2$,
  regardless of the topology of the underlying network. Results for 
  uncorrelated scale-free networks having degree distribution $P(k)\sim
  k^{-\gamma}$ with $\gamma=2.5$ and $\gamma=3.0$ and different sizes $V$ 
show the same
  behavior, small differences in the value of $\rho_B$ being due to finite-size effects.  If $D_A=1$, 
the critical point is strongly affected by the
  topological fluctuations of the network.  Here we show results for
  $\gamma=2.5$ and sizes of the network $V=10^3,\,10^4,\,10^5$
  corresponding to $\la k\ra^2/\la k^2\ra =0.52,\,0.32,\,0.19$, respectively. 
With increasing sizes, degree
  fluctuations become larger and the transition is observed at smaller
  values of $\rho_c$. Bottom panels: Stationary densities $\rho_{A,k}$
  and $\rho_{B,k}$ as functions of the degree $k$. If $D_A=0$, 
the average density of $A$ particles inside nodes of
  degree $k$ is constant, while the behavior shown by $B$ particles is
  linear in $k$. If $D_A=1$, both
  densities are linear in $k$.\\

\noindent
{\bf Figure 3.} {\bf Phase diagram and stationary densities for type II
    processes.} Top panel: Phase transitions in type II processes for
  diffusing and non-diffusing $A$ particles. Regardless of topological
  fluctuations in the underlying network and of the probability of
  diffusion $D_A$, the transition occurs at the critical point
  $\beta/\mu=1$, depending only on the reaction rates. Here we show
  results for networks of size $V=10^4$ with particle density
  $\rho=20$, power-law exponents $\gamma=2.5$ and $\gamma=3.0$.
  Differences in the values of the stationary density $\rho_B$ are due
  to finite-size effects. Bottom panels: Stationary densities
  $\rho_{A,k}$ and $\rho_{B,k}$ as functions of the degree $k$. In both
  cases, $D_A=0$ and $D_A=1$, linear dependencies in $k$ are
  obtained.\\

\noindent
{\bf Figure 4.} {\bf Reaction activity in type I and type II processes: microscopic model and real-world examples.}  
The first row of panels refers to the microscopic
RD model, as described in the text, while the second row reports the analysis of 
the spread of an airline-carried disease in the US with a data-driven meta-population model. Both models consider the actual topology of the US air transportation network as obtained by considering the 500 airports with largest traffic~\cite{iata}; nodes population is obtained from census data~\cite{census}. In addition,
the realistic meta-population model also considers the actual traffic of passengers on each connection between airports.
The networks are mapped on a globe
for sake of visualization. Each node is represented
with a size linearly dependent on its population and a color illustrating
the level of  reaction activity inside the node, ranging from 0 to the max 
value.
Whereas type I processes experience a 
level of  activity proportional to nodes population~--~corresponding to  red color
in largely populated nodes and  yellow in small population nodes~--~the  reaction activity is homogeneously distributed among the nodes of the network when type II processes are considered.

\newpage

\begin{figure}[!ht]
\begin{center}
\includegraphics[width=10cm]{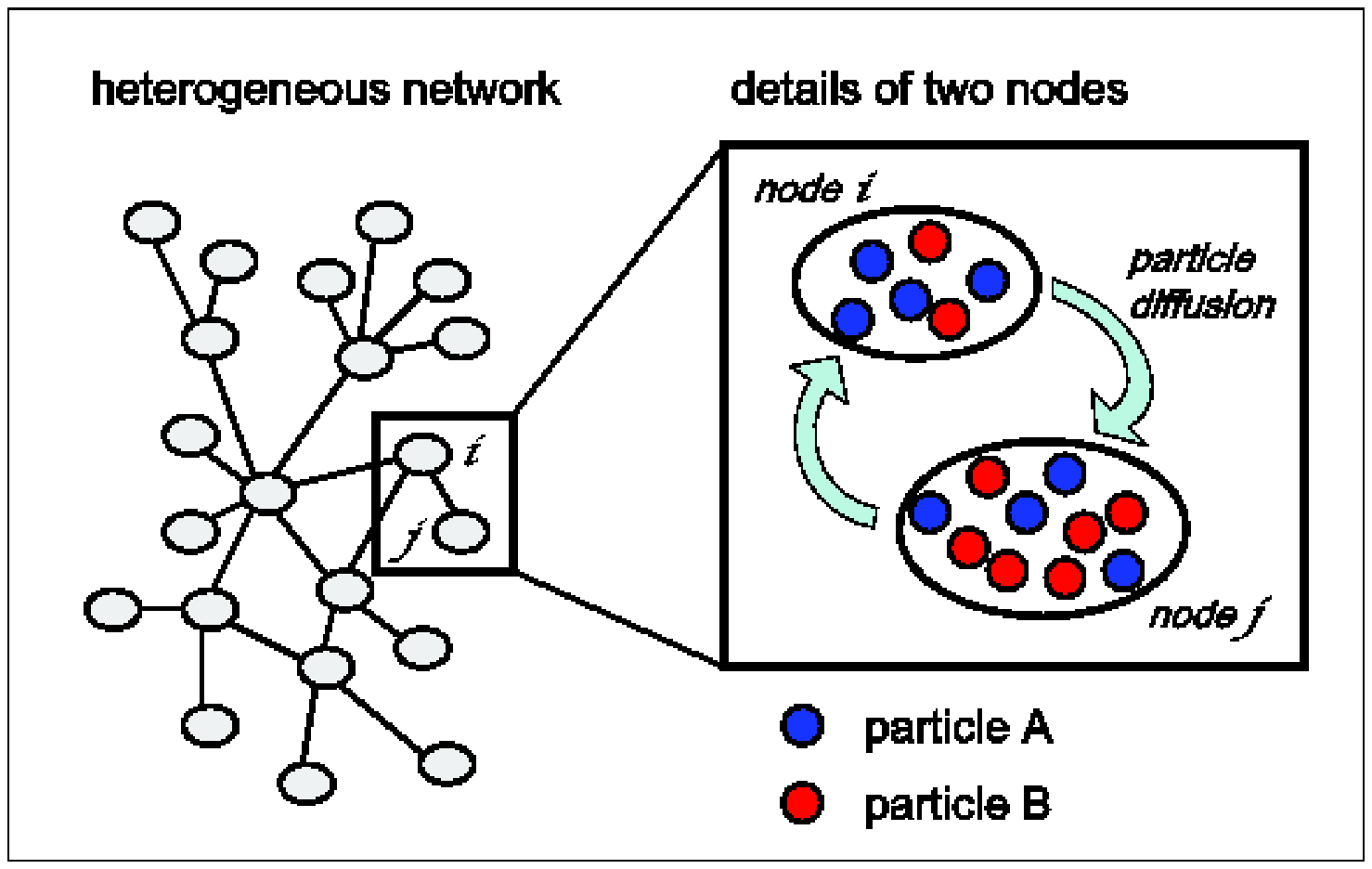}

{\bf Figure 1.}

\end{center}
\end{figure}

\begin{figure}[!ht]
\begin{center}
\vskip .7cm
\includegraphics[width=12cm]{fig2}

{\bf Figure 2.}

\end{center}
\end{figure}

\begin{figure}[!ht]
\begin{center}
\includegraphics[width=12cm]{fig3}

{\bf Figure 3.}

\end{center}
\end{figure}

\begin{figure}[!ht]
\begin{center}
\includegraphics[width=18cm]{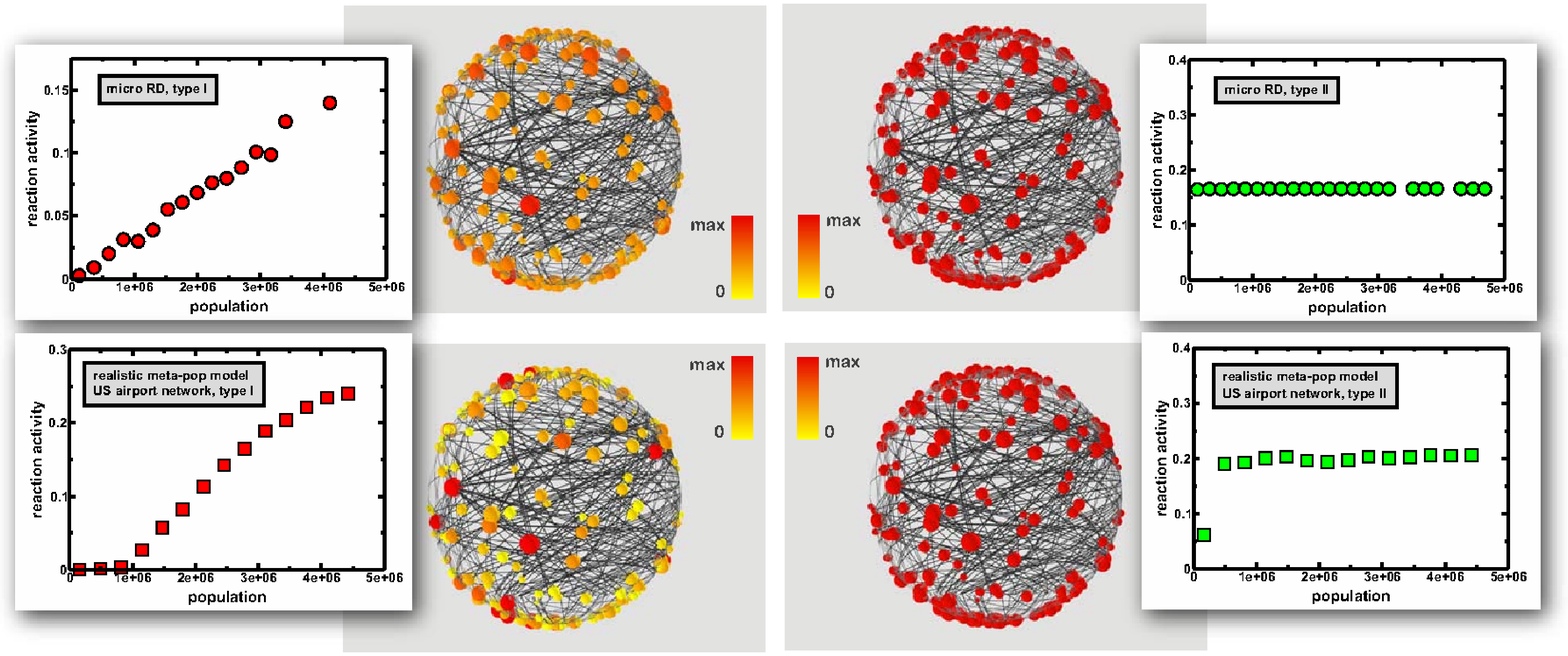}

{\bf Figure 4.}

\end{center}
\end{figure}

\end{document}